\documentclass[12pt]{article}
\usepackage{amsmath, amssymb}
\usepackage{graphicx,psfrag, subcaption}
\usepackage{enumerate}
\usepackage{natbib}
\usepackage{bbm}
\usepackage{url} 

\newcommand{\blind}{0}

\DeclareMathOperator{\nach}{ne}

\DeclareMathOperator{\T}{^{\mathsf T}}
\def\V{\mathcal{V}}
\def\E{\mathcal{E}}
\def\G{\mathcal{G}}

\newcommand{\indep}{\mathrel{{\perp}\hspace*{-0.6em}{\perp}}}
\newcommand{\given}{\mathrel{|}}
\newcommand{\condindep}[3]{#1 \indep #2 \given #3}
\newtheorem{definition}{Definition}

\begin{document}

\def\spacingset#1{\renewcommand{\baselinestretch}%
{#1}\small\normalsize} \spacingset{1}


\if0\blind
{
  \title{\bf  Graphical modelling of multivariate spatial point processes}
  \author{Matthias Eckardt\hspace{.2cm}\\
    Department of Computer Science, Humboldt Universit\"{a}t zu Berlin, Berlin, Germany}
  \maketitle
} \fi

\if1\blind
{
  \bigskip
  \bigskip
  \bigskip
  \begin{center}
    {\LARGE\bf Title}
\end{center}
  \medskip
} \fi

\bigskip
\begin{abstract}
This paper proposes a novel graphical model, termed the spatial dependence graph model, which captures the global dependence structure of different events that occur randomly in space. In the spatial dependence graph model, the edge set is identified by  using the conditional partial spectral coherence. Thereby, nodes are related to the components of a multivariate spatial point process and edges express orthogonality relation between the single components. This paper introduces an efficient approach towards pattern analysis of highly structured and high dimensional spatial point processes. Unlike all previous methods, our new model permits the simultaneous analysis of all multivariate conditional interrelations. The potential of our new technique to investigate multivariate structural relations is illustrated using data on forest stands in Lansing Woods as well as monthly data on crimes committed in the City of London.  
\end{abstract}

\noindent%
{\it Keywords:} Crime data, Dependence structure; Graphical model; Lansing Woods, Multivariate spatial point pattern
\vfill

\section{Introduction}

The analysis of spatial point patterns is a rapidly developing field and of particular interest to many disciplines. Here, a main concern is to explore the structures and relations generated by a countable set of randomly occurring points in some bounded planar observation window. Generally, these randomly occurring points could be of one type (univariate) or of two and more types (multivariate). In this paper, we consider the latter type of spatial point patterns. 

Recently, multivariate spatial point patterns have gained much attraction and applications have been published in various fields including epidemiology \citep{Diggle2005}, ecology \citep{Illian2007, Shimatani2001} and forestry \citep{Grabarnik2009}. Although various well-established methods exist, mostly uni- or bivariate, the need for efficient exploratory techniques for multivariate spatial point patterns still remains. Of particular note, the challenge in detecting global conditional structural relations that might exist with respect to all components of a multivariate spatial point pattern has not been approached.

In order to overcome this limitation, we introduce a novel graphical model, termed the spatial dependence graph model (SDGM).  In general, graphical models combine probability theory and graph theory and thus provide a suitable approach of dealing with uncertainty and complexity using conditional independence statements and factorisations of joint densities. Thus, they graphically display Markov properties most commonly related to random variables. For a profound treatment of different graphical models the interested reader is referred to  \citet{Cowell1999, Cox1996, Edwards2000,Koller2010,  Lauritzen1996, Pearl1988, Spirtes2000} and \citet{ Whittaker2008}.

In recent years, a growing number of temporal extensions of various graphical models have been presented. For point process data, these models include the local dependence graph model \citep{Didelez2000, Didelez2007, Didelez2008}, the dynamic path analysis model \citep{Aalen2008, Fosen2006, Martinussen2010} as well as the graphical duration model \citep{Dreassi2007,Gottard2007,Gottard2007a} besides others. Alternatively, \citet{Brillinger1996,Dahlhausetal1997} and \citet{Eichler2003} discussed graphical models for point process data in the frequency domain. SDGMs can be seen as extension of these frequency domain models for spatial point patterns. A comprehensive review of temporal graphical models is given in \citet{Eckardt:2016}. 

In a SDGM, the global conditional interrelation structure between components of possibly highly complex as well as high dimensional multivariate spatial point patterns is depicted by means of an undirected graph. In the spatial domain, undirected graphs have most commonly been used to capture spatial neighbourhood relations of lattice data. For spatial point patterns, undirected graphs have only been considered for neighbour networks (\citet{Marchette2004, Penrose2003,Penrose2005,Penrose2001}). Different to SDGMs, neighbour networks are random graph models where single events are represented by distinct nodes, while undirected graphs have appeared as a new spatial domain within the class of point processes on linear networks (\citet{Ang2010, Ang2012,Baddeley2014, Okabe2001}). The idea of analysing the global conditional interrelation structure between components of multivariate spatial point patterns using a graph model is new. 

This paper is structured as follows: Section \ref{sec:1} presents the basic graph theoretical concepts, discusses the properties of spatial point processes in the spatial and frequency domain and  defines the SDGM. Applications of the SDGM to two real data sets are given in Section \ref{sec:2}.  Finally, the theoretical results are discussed in Section \ref{sec:3}.

\section{Graphical modelling of spatial point patterns}\label{sec:1}

This section introduces a new framework for the analysis of global conditional interrelations in $d$-variate spatial point patterns where we assume $d\geq 3$. To this end, we define a novel graphical model whose graphical structure is related to conditional spectral properties of a finite set of randomly occurring points of different types in a bounded region. Hence, we relate the dependence structure of a multivariate point process to the adjacency structure encoded in a graph. Before discussing spatial point patterns and defining SDGMs, some basic notation and terminology on point processes and graph theory is needed. Firstly, we introduce the notation for point processes. For a rigorous discussion of the theoretical fundamentals of point processes the interested reader is referred to  \citet{Cox1980} and also \citet{Daley2003, Daley2008}. Basically, let $\mathbf{Z}=(Z_1,\ldots, Z_d)$ denote a $d$-variate spatial point process generating $d$ different types of points which we call events. The realisations of this multivariate point process can then be represented as $d$-variate counting process $\mathbf{N}_\V$ with index set $\V=\lbrace1,\ldots, d\rbrace$ and components $N_i,~i\in\V$. Here, $N_i(\alpha)$ counts the number of events of type $i$ appearing in an arbitrary region $\alpha \subset \mathbbm{R}^2$. Additionally, we write $N_A=(N_i)_{i\in A}$ to refer to a subprocess of $\mathbf{N}_\V$.
 
Next, we introduce some notation and terminology of graph theory needed for discussion. An in-depth treatment of graph theory is given in \citet{Bondy2008} and also \citet{Diestel2010}. A graph is defined as pair $\G=(\V,\E)$ where $\V=\left\{v_1,\dots,v_k\right\}$ is a finite set of vertices or nodes and $\E\subseteq\V\times\V$ is a finite set of edges -- joining the vertices where $\E(\G)\cap\V(\G)=\emptyset$. Throughout this paper, we only consider undirected graphs, that is we only allow for undirected edges. If two nodes are joined by an edge they are called adjacent and the set of all adjacent nodes of a distinct node $v_j$ is the neighbourhood $\nach\left(v_j\right)=\lbrace
v_i: (v_i,v_j)\in\E(\G)\rbrace$. A sequence of potentially repeating vertices and edges $\left(v_0,e_1,v_1,e_2,\ldots,v_{k-1},e_k,v_k\right)$ in $\G$ with endpoints $v_0$ and $v_k$ such that $\forall~e_i, 1\leq i\leq k $ is a walk of length $k$. If a walk passes through every node of a sequence exactly once we label this as a path. If every distinct pair of vertices in $\G$ is joined by a path, $\G$ is said to be connected. A component is a non-empty maximal connected subgraph $\G'$ of $\G$ such that every distinct pair of nodes is joined by a path in $\G'$. Lastly, suppose a partition of the vertices such that $v_i, v_j\in\V(G)$ and $\mathcal{S}\subset\V(\G)\backslash\lbrace v_i,v_j\rbrace$. Then, $\mathcal{S}$ is a separating set or ij-vertex-cut of $\G$ if and only if $v_i$ and $v_j$ are not in the same component in $\G\setminus\mathcal{S}$.

\subsection{Properties of spatial point patterns}\label{spprob}

Before we discuss the spectral properties of spatial point processes and define the SDGM, we first need to introduce statistical measures for describing the first and second-order properties of spatial point processes in the spatial domain. The first-order properties are related to the mean number of events per unit area while the second-order properties express the variance of the number of events per unit area. A profound treatment on the statistical analysis of spatial point processes is given in \citet{Diggle2002, Illian2008} and \citet{Moller2004}.  

Usually, the first-order properties of a spatial point process are expressed by means of the first-order intensity function. Following the notation of \citet{Diggle2002, Diggle2013}, the first-order intensity function is given as 
\[
\lambda_i(\mathbf{s})=\lim_{|d\mathbf{s}|\rightarrow 0}\left\{\frac{\mathbbm{E}\left[ N_i(d\mathbf{s}))\right]}{|d\mathbf{s}|}\right\}, \mathbf{s}\in\mathbf{S}.
\]  
Here, $\mathbf{s}=(x, y)$ is the location of a randomly occurring event within a bounded region $\mathbf{S}\subset\mathbbm{R}^2$, $N_i(d\mathbf{s}) = N_i(\mathbf{s}+d\mathbf{s}) - N_i(\mathbf{s})$ is the number of observed events of type $i$ within a infinitesimal region containing $\mathbf{s}$ and $|d\mathbf{s}|$ denotes the area of $d\mathbf{s}$. 

To describe the second-order properties of a spatial point pattern, one can use the second-order intensity function. For any pair of locations $\mathbf{s}=(x,y)$ and $\mathbf{s'}=(x',y')$, the second-order intensity function is defined as
\[
\lambda_{ii}(\mathbf{s,s'})=\lim_{|d\mathbf{s}|,|d\mathbf{s}|\rightarrow 0}\left\{\frac{\mathbbm{E}\left[N_i(d\mathbf{s})N_i(d\mathbf{s'})\right]}{|d\mathbf{s}||d\mathbf{s'|}}\right\}, \mathbf{s}\neq\mathbf{s'}, \mathbf{s},\mathbf{s'}\in\mathbf{S}.
\]

Although the second-order intensity function is connected to the highly prominent reduced second-order moment function also known as Ripleys' $K$-function \citep{Ripley1976}, it is less useful to describe the theoretical properties of spatial point patterns. A more suitable function in this respect is the covariance density function. Here, focussing on multivariate spatial point patterns, we concern the auto-covariance and the cross-covariance density function which express the second-order properties within and between component processes. The auto-covariance density function is defined as
\begin{equation}
\gamma_{ii}(\mathbf{s,s'})=\lim_{|d\mathbf{s}|,|d\mathbf{s'}|\rightarrow 0}\left\{\frac{\mathbbm{E}\left[\lbrace N_i(d\mathbf{s})-\lambda_i(d\mathbf{s})\rbrace \lbrace N_i(d\mathbf{s'})-\lambda_i(d\mathbf{s'})\rbrace\right]}{|d\mathbf{s}||d\mathbf{s'|}}\right\}
\label{autocov}
\end{equation}
and can be obtained from the second-order intensity function as
\[
\gamma_{ii}(\mathbf{s,s'})=\lambda_{ii}(\mathbf{s,s'})-\lambda_i(\mathbf{s})\lambda_i(\mathbf{s'}).
\]
In addition, the cross-covariance density function for any two disjoint events $i$ and $j$ follows as
\begin{equation}
\gamma_{ij}(\mathbf{s,s'})=\lim_{|d\mathbf{s}|,|d\mathbf{s'}|\rightarrow 0}\left\{\frac{\mathbbm{E}\left[\lbrace N_i(d\mathbf{s})-\lambda_i(d\mathbf{s})\rbrace\lbrace N_j(d\mathbf{s'})-\lambda_j(d\mathbf{s'})\rbrace\right]}{|d\mathbf{s}||d\mathbf{s'|}}\right\}.
\label{crosscov}
\end{equation}

For orderly processes, which imply that only one event can occur at a particular location,  \eqref{autocov} and \eqref{crosscov} include the case when $\mathbf{s}=\mathbf{s'}$. Precisely, for orderly processes we have $
\mathbbm{E}\left[\lbrace N_i(d\mathbf{s})\rbrace^2\right]=\lambda_i(\mathbf{s})|d\mathbf{s}|$. The integration of this expression into the covariance density function leads to Bartletts' complete auto-covariance density function $\kappa_{ii}(\cdot)$  \citep{Bartlett1964}, namely
\begin{equation}\label{completecov}
\kappa_{ii}(\mathbf{s,s'})=\lambda_i(\mathbf{s})\delta(\mathbf{s}-\mathbf{s'})+\gamma_{ii}(\mathbf{s,s'})
\end{equation}
where $\delta(\cdot)$ denotes a two-dimensional Dirac delta function.

Generalisations of orderliness to $d$-variate processes follow naturally such that 
\[
\mathbbm{E}\left[N_i(d\mathbf{s}),\ldots, N_d(d\mathbf{s}))\right]=0.
\]
Similar to the covariance density function, we consider the complete auto-covariance and the complete cross-covariance density function. As proposed in \cite{Mugglestone1996a}, we define the complete cross-covariance for events of types $i$ and $j$ as $\kappa_{ij}(\mathbf{s,s'})=\gamma_{ij}(\mathbf{s,s'})$ and $\kappa_{ji}(\mathbf{s,s'})=\gamma_{ji}(\mathbf{s,s'})$.

\subsection{Spectral properties of spatial point patterns}\label{sec:spectralproperties}

Fourier transformations and spectral analysis techniques determine the presence of periodic structures in spatial point processes and present a complementary approach to distance-related methods. Different from inter-distance techniques and statistical models in the spatial domain, spectral analysis methods do not require any prior distributional assumptions and allow for anisotropic or non-stationary processes and also different scales. 

Different to the analysis of time series data, spectral techniques have not been studied and applied with respect to spatial point processes much so far. Thus, the number of methodological and applied contributions remain limited although certain advantages exist. The spectral analysis of temporal point processes has first been considered by \citet{Bartlett1963} and also by \citet{Brillinger1972}. \citet{Bartlett1964} presented an extension of his work to two-dimensional point processes. A profound treatment of spectral properties with respect to spatial point processes is given in \citet{Renshaw1997, Renshaw2002, Renshaw1983, Renshaw1984} and \citet{Mugglestone1996a,Mugglestone1996b,  Mugglestone2001}. 

In order to discuss the spectral properties theoretically, we assume the spatial point process to be orderly - such that multiple coincident events can not occur - and second-order stationary. Second-order stationarity implies, that the first-order intensity function $\lambda_i(\mathbf{s}), \mathbf{s}\in\mathbf{S}$ is constant over a finite region $\mathbf{S}\subset \mathbbm{R}^2$ while the covariance density function $\gamma_{ij}(\mathbf{s,s'})$ depends on $\mathbf{s}$ and $\mathbf{s'}$ only through $\mathbf{c}=\mathbf{s}-\mathbf{s'}$. For a $d$-variate process the notion of stationarity implies that all $d$ processes are marginally and jointly stationary. Consequently, we have $\gamma_{ii}(\mathbf{s,s'})=\gamma_{ii}(\mathbf{c})$ and also $\kappa_{ii}(\mathbf{s,s'})=\kappa_{ii}(\mathbf{c})$. For the covariance density function we notice that $\gamma_{ij}(\mathbf{s,s'})=\gamma_{ji}(\mathbf{s',s})$ such that stationarity also implies that $\gamma_{ij}(\mathbf{c})=\gamma_{ji}(-\mathbf{c})$ and $\kappa_{ij}(\mathbf{c})=\kappa_{ji}(-\mathbf{c})$ (cf. \citet{Mugglestone1996a, Mugglestone1996b}).   

For a second-order stationary spatial point process the auto-spectral density function for event $i$ at frequencies $\boldsymbol{\omega}=(\omega_1,\omega_2)$ appears as the Fourier transform of the complete auto-covariance density function of $N_i$,  
 \begin{equation}\label{fouriereq}
\begin{split}
f_{ii}(\boldsymbol{\omega}) &= \int \kappa_{ii}(\mathbf{c})\exp(-\iota\boldsymbol{\omega}^{\T}\mathbf{c})d\mathbf{c}\\
&=\int^\infty_{-\infty}\int^\infty_{-\infty}\kappa_{ii}(c_1,c_2)\exp\lbrace-\iota(\omega_1c_1+\omega_2c_2)\rbrace dc_1dc_2
\end{split}
\end{equation}
where $\iota=\sqrt{-1}$ and $\omega^{\T}$ denotes the transpose of $\mathbf{\omega}$. As described in \citet{Brillinger1981} and \cite{Brockwell2006} with respect to time series, the auto-spectrum can be understood as decomposition of $\kappa_{ii}$ into a periodic function of frequencies $\boldsymbol{\omega}$.

From expression \eqref{fouriereq}, the complete auto-covariance density function can uniquely be recovered via inverse Fourier transformation,
\begin{equation}\label{inversekappa}
\kappa_{ii}(\mathbf{c}) = \int f_{ii}(\boldsymbol{\omega})
\exp\left(\iota\boldsymbol{\omega}^{\T}\mathbf{c}\right)d\boldsymbol{\omega}.
\end{equation}

Substituting for $\kappa_{ii}(\mathbf{c})$ from \eqref{completecov} finally leads to 
 \begin{equation}\label{fouriereqfinal}
f_{ii}(\boldsymbol{\omega}) =\lambda_i+\int^\infty_{-\infty}\int^\infty_{-\infty}\gamma_{ii}(c_1,c_2)\exp\lbrace-\iota(\omega_1c_1+\omega_2c_2)\rbrace dc_1dc_2.
\end{equation}

Similarly, the cross-spectral density function is given as Fourier transform of the complete cross-covariance density function,
\begin{equation}\label{crossspectrakappa}
f_{ij}(\boldsymbol{\omega}) = \int \kappa_{ij}(\mathbf{c})\exp(-\iota\boldsymbol{\omega}^{\T}\mathbf{c})d\mathbf{c},
\end{equation}
which measures the linear interrelation of components $N_i$ and $N_j$. Thus, two processes are said to be uncorrelated at all spatial lags if and only if the corresponding spectrum is zero at all frequencies. In addition, since $\kappa_{ij}(\mathbf{c})=\kappa_{ji}(-\mathbf{c})$ we equivalently have $f_{ij}(\mathbf{c})=f_{ji}(-\mathbf{c})$ and thus it is sufficient to calculate only one cross-spectrum (cf. \citet{Bartlett1964, Mugglestone1996a, Mugglestone1996b}). 

Usually, the cross-covariance function could be asymmetric, namely $\gamma_{ij}(\mathbf{c})\neq \gamma_{ji}(\mathbf{-c})$, such that the cross-spectrum is a complex-valued function. As discussed in \citet{Chatfield1989} and \citet{Priestley1981}, a common procedure in time series analysis is to split the complex-valued cross-spectrum into the real and the imaginary part, namely into the co-spectrum $C_{ij}(\boldsymbol{\omega})$ and quadrature spectrum $Q_{ij}(\boldsymbol{\omega})$ at frequencies $\boldsymbol{\omega}$. Thus, the cross-spectrum can be decomposed in terms of Cartesian coordinates as
\[
\begin{split}
f_{ij}(\boldsymbol{\omega})&=\frac{1}{2\pi}\sum^\infty_{\mathbf{c}=-\infty}\cos(\boldsymbol{\omega}^{\T}\mathbf{c})\kappa_{ij}(\mathbf{c})-\iota \frac{1}{2\pi}\sum^\infty_{\mathbf{c}=-\infty}\sin(\boldsymbol{\omega}^{\T}\mathbf{c})\kappa_{ij}(\mathbf{c})\\
&=C_{ij}(\boldsymbol{\omega})-\iota Q_{ij}(\boldsymbol{\omega}).
\end{split}
\]
Alternatively, using polar coordinates, the cross-spectrum can be expressed in terms of its modulus and its phase as $f_{ij}(\boldsymbol{\omega})=\zeta_{ij}(\boldsymbol{\omega})\exp\lbrace \iota\varphi_{ij}(\boldsymbol{\omega})\rbrace$. Here,   
\begin{equation}\label{Eq:crossamplitude}
\begin{split}
\zeta_{ij}(\boldsymbol{\omega})&=\mod\lbrace f_{ij}(\boldsymbol{\omega})\rbrace\\
&=\sqrt{\left(\lbrace C_{ij}(\boldsymbol{\omega})\rbrace^2+\lbrace Q_{ij}(\boldsymbol{\omega})\rbrace^2\right)}
\end{split}
\end{equation}
is called the cross-amplitude spectrum and measures the relative magnitude of the power attributable to  frequencies $\boldsymbol{\omega}$ in a bivariate point pattern. The second term,   
\begin{equation}\label{Eg:cross.Phase}
\begin{split}
\varphi_{ij}(\boldsymbol{\omega})&=\arg\lbrace f_{ij}(\boldsymbol{\omega})\rbrace\\
=&\tan^{-1}\left\{\frac{-Q_{ij}(\boldsymbol{\omega})}{C_{ij}(\boldsymbol{\omega})}\right\},
\end{split}
\end{equation}
is called the cross-phase spectrum and indicates how closely linear translations of the pattern formed by one component match the pattern formed by the other component. Hence, $\varphi_{ij}(\boldsymbol{\omega})$ measures the similarity of two patterns up to linear shifts (cf. \citet{ Chatfield1989, Priestley1981}). This information is provided by the slope of the phase which measures the magnitude and direction of the shift. Obviously, the phase is undefined whenever the cross-spectrum vanishes and its meaning is questionable if only small values of the cross-spectrum appears.

Although the cross-spectrum expresses the linear interrelation between two component processes, it is often preferable to use the spectral coherence as rescaled version of the cross-spectrum. The spectral coherence is defined as
\begin{equation}\label{eq:Coh}
\vert R_{ij}(\boldsymbol{\omega})\vert^2=\frac{f_{ij}(\boldsymbol{\omega})^2}{\left[f_{ii}(\boldsymbol{\omega})f_{jj}(\boldsymbol{\omega})\right]}
\end{equation}
and measures the linear relation of two components. Different from the auto-spectrum resp. cross-spectrum we have that $0\leq \vert R_{ij}(\boldsymbol{\omega})\vert^2\leq 1$.

However, the spectral coherence is not able to distinguish between direct and induced interrelations. In order to control for the linear effect of all remaining component processes $N_{V\backslash\lbrace i,j\rbrace}$ on pairwise linear interrelations between $N_i$ and $N_j$, we adopt the framework of partialisation. Thus, in analogy with graphical modelling of multivariate data, we are interested in the linear interrelation between $N_i$ and $N_j$ that remains after elimination of the linear effect of all alternative component processes. In this respect, the partial cross-spectrum $f_{ij\given\V\backslash\lbrace i,j\rbrace}(\boldsymbol{\omega})$ follows as cross-spectrum of the residual processes $\epsilon_i$ and $\epsilon_j$ which result from the elimination of the linear effect of $N_{V\backslash\lbrace i,j\rbrace}$ on $N_i$ and $N_j$. So, we have $f_{ij\given\V\backslash\lbrace i,j\rbrace}(\boldsymbol{\omega})=f_{\epsilon_i\epsilon_j}(\boldsymbol{\omega}).$

With respect to the calculation of the partial cross-spectrum, different methods have been proposed in the literature. Adopting the results of \citet[Theorem 8.3.1.]{Brillinger1981}, we can compute the partial cross-spectrum using the formula
\begin{equation}\label{partial.formula}
f_{ij\given\V\backslash\lbrace i,j\rbrace}(\boldsymbol{\omega})=f_{ij}(\boldsymbol{\omega})-f_{i\V\backslash\lbrace i,j\rbrace}(\boldsymbol{\omega})f_{\V\backslash\lbrace i,j\rbrace\V\backslash\lbrace i,j\rbrace}(\boldsymbol{\omega})^{-1}f_{\V\backslash\lbrace i,j\rbrace j}(\boldsymbol{\omega})
\end{equation}
where \[
f_{i\V\backslash\lbrace i,j\rbrace}(\boldsymbol{\omega})=\left[f_{i1}(\boldsymbol{\omega}), \ldots,f_{ii-1}(\boldsymbol{\omega}), f_{ii+1}(\boldsymbol{\omega}),\ldots,f_{ij-1}(\boldsymbol{\omega}), f_{ij+1}(\boldsymbol{\omega}), \ldots, f_{ik}(\boldsymbol{\omega}) \right].
\]
Obviously, the computation of \eqref{partial.formula} requires the inversion of a $(d-2)\times(d-2)$ matrix. As an alternative approach, one can implement a step-wise procedure for \eqref{partial.formula} by recursively applying algebraic operations as described in \citet{Bendat1978}. 

To illustrate this recursive calculus, we consider a five-dimensional counting process with components $N_i$ to $N_m$. Here, we are interested in the partial cross-spectrum of order three, e.g. $f_{ij|klm}(\boldsymbol{\omega})$. In the initial step, the partial cross-spectrum of order one, e.g. $f_{ij|k}(\boldsymbol{\omega})$, results from the equation
\begin{equation}\label{firstorderpartial}
f_{ij|k}(\boldsymbol{\omega}) =f_{ij}(\boldsymbol{\omega})-f_{ik}(\boldsymbol{\omega})f_{kk}(\boldsymbol{\omega})^{-1}f_{kj}(\boldsymbol{\omega}).
\end{equation}
Similarly, all remaining partial spectra of order one can be obtained by replacing the corresponding spectral functions of the right-hand-side of \eqref{firstorderpartial}. 
To obtain the partial spectra of order two, all auto-spectra and cross-spectra terms of \eqref{firstorderpartial} are recursively replaced by their partial spectra counterparts of order one.
Consequently, for $f_{ij|kl}(\boldsymbol{\omega})$ we have 
\[
f_{ij|kl}(\boldsymbol{\omega}) =  f_{ij|k}(\boldsymbol{\omega})-f_{il|k}(\boldsymbol{\omega})f_{ll|k}(\boldsymbol{\omega})^{-1}f_{lj|k}(\boldsymbol{\omega}).
\]
Hence, the partial spectra of order $k$ are calculated by substituting the corresponding partial auto-spectra and partial cross-spectra expressions of order $k-1$ for the right-hand side terms of \eqref{firstorderpartial}.     
Thus, we achieve the desired partial cross-spectra of order three as 
\[
f_{ij|klm}(\boldsymbol{\omega})=f_{ij|kl}(\boldsymbol{\omega})-f_{im|kl}(\boldsymbol{\omega})f_{mm|kl}(\boldsymbol{\omega})^{-1}f_{mj|kl}(\boldsymbol{\omega}).
\]

However, this recursive procedure is less computationally efficient in case of high dimensional processes as the required number of stepwise calculations depends on the number of distinct component processes. 

A less computationally intensive approach has recently been introduced in \cite{Dahlhaus2000} where, under regularity assumptions, the partial spectra can be obtained from the inverse of the spectral matrix, namely $\mathbf{g}(\boldsymbol{\omega})=\mathbf{f}(\boldsymbol{\omega})^{-1}$. Different to \eqref{partial.formula}, this approach only requires the inversion of a $d\times d$ matrix.

Analogously to \eqref{eq:Coh}, the partial spectral coherences is obtained as rescaled version of the partial cross-spectrum, 
\begin{equation}
|R_{ij\given\V\backslash\lbrace i,j\rbrace}(\boldsymbol{\omega})|^2=\frac{f_{ij\given\V\backslash\lbrace i,j\rbrace}(\boldsymbol{\omega})^2}{\left[f_{ii\given\V\backslash\lbrace i,j\rbrace}(\boldsymbol{\omega})f_{jj\given\V\backslash\lbrace i,j\rbrace}(\boldsymbol{\omega})\right]}.
\end{equation}

Alternatively, applying the results of \citet[Theorem 2.4.]{Dahlhaus2000}, we can efficiently compute the partial spectral coherence $|R_{ij\given\V\backslash\lbrace i,j\rbrace}(\boldsymbol{\omega})|^2$ from $g_{ij}(\boldsymbol{\omega})$, where
\begin{equation}
\label{InverseSpectraRxyz}
R_{ij\given\V\backslash\lbrace i,j\rbrace}(\boldsymbol{\omega})=-\frac{g_{ij}(\boldsymbol{\omega})}{\left[ g_{ii}(\boldsymbol{\omega})g_{jj}(\boldsymbol{\omega})\right]^{\frac{1}{2}}}.
\end{equation}
More precisely, under regularity assumptions, we can define the absolute rescaled inverse as
\begin{equation}
\vert d_{ij}(\boldsymbol{\omega})\vert=\frac{\vert g_{ij}(\boldsymbol{\omega}) \vert}{\left[g_{ii}(\boldsymbol{\omega})g_{jj}(\boldsymbol{\omega})\right]^{\frac{1}{2}}}
\end{equation}
which measures the strength of the linear partial interrelation between $N_i$ and $N_j$ at frequencies $\boldsymbol{\omega}$. As shown in \cite{Dahlhaus2000}, we then have
\begin{equation}
d_{ij}(\boldsymbol{\omega})=-R_{ij\given\V\backslash{\lbrace i,j\rbrace}}(\boldsymbol{\omega})
\end{equation} 
such that we can obtain the partial spectral coherence from the negative of the absolute rescaled inverse.

As for the ordinary spectral coherence, we also have  $0\leq\vert R_{ij\given\V\backslash\lbrace i,j\rbrace}(\mathbf{\omega})\vert^2\leq 1$. Here, $\vert R_{ij\given\V\backslash\lbrace i,j\rbrace}(\mathbf{\omega})\vert^2=0$ indicates conditional orthogonality of $N_i$ and $N_j$ given all remaining component processes ($\condindep{N_i}{N_j}{N_{\V\backslash\lbrace i,j\rbrace}}$) while $\vert R_{ij\given\V\backslash\lbrace i,j\rbrace}(\mathbf{\omega})\vert^2=1$ expresses a perfect linear relation of $N_i$ and $N_j$ given $N_{\V\backslash\lbrace i,j\rbrace}$.

 Thus different from the ordinary spectral coherence, the partial spectral coherence expresses the linear interrelation of two component processes which remains after the linear effect of all remaining component processes has been removed by orthogonal projection. In this sense, the partial spectral coherence can be understood as the partial correlation defined as a function of frequencies $\boldsymbol{\omega}$ \citep{Brillinger1981,Rosenberg1989}. 

As shown in \cite{Eichler2003}, we also have the following relations between the partial auto-spectrum, partial cross-spectrum and the partial spectral coherence which result from  the inverse variance lemma \citep{Whittaker2008}: 
\[
f_{ii\given\V\backslash\lbrace i,j\rbrace}(\boldsymbol{\omega})=\frac{f_{ii\given\V\backslash\lbrace i\rbrace}(\boldsymbol{\omega})}{1-|R_{ij\given\V\backslash\lbrace i,j\rbrace}(\boldsymbol{\omega})|^2}
\]
and 
\[
f_{ij\given\V\backslash\lbrace i,j\rbrace}(\boldsymbol{\omega})=\frac{R_{ij\given\V\backslash\lbrace i,j\rbrace(\boldsymbol{\omega})}(\boldsymbol{\omega})}{1-|R_{ij\given\V\backslash\lbrace i,j\rbrace}(\boldsymbol{\omega})|2}\times\sqrt{f_{ii\given\V\backslash\lbrace i\rbrace}(\boldsymbol{\omega})f_{jj\given\V\backslash\lbrace j\rbrace}(\boldsymbol{\omega})}.
\]
 
All previous orthogonality statements also hold for disjoint subprocesses of $\mathbf{N}_\V$. So, for $\lbrace N_I: i\in I\subset N_\V\rbrace, \lbrace N_J: j\in J\subset N_\V\rbrace$ and $\lbrace N_H: h\in H\subset N_\V\rbrace$ we have  
\begin{eqnarray*}
\condindep{N_I}{N_J}{N_H} &\Leftrightarrow  f_{IJ\given H}(\boldsymbol{\omega})=0\\
  &\Leftrightarrow R_{IJ\given H}(\boldsymbol{\omega})=0.
\end{eqnarray*}

\subsection{Isotropic processes}

Further simplifications of the auto-spectral and cross-spectral density functions appear if the probabilistic statements about a process are isotropic, namely invariant under rotation of $\mathbf{S}$. Isotropy implies that the second-order properties for locations $\mathbf{s}$ and $\mathbf{s'}$ only depend on the scalar distance between $\mathbf{s}$ and $\mathbf{s'}$. Thus, for a stationary and isotropic process we have $\gamma_{ij}(\mathbf{c})=\gamma_{ij}(\Vert\mathbf{c}\Vert)$ and $\kappa_{ij}(\mathbf{c})=\kappa_{ij}(\Vert\mathbf{c}\Vert)$ where $\Vert\mathbf{c}\Vert=\sqrt{c_1^2+c_2^2}$. Here, we observe that for a stationary and isotropic process $\gamma_{ij}(\Vert\mathbf{c}\Vert)=\gamma_{ji}(\Vert\mathbf{c}\Vert)$ such that it suffices to take only one covariance density function into account (cf. \citet{Diggle2002}). A simplification of the auto-spectral density function can then be achieved by expressing $\mathbf{c}$ in terms of polar coordinates, namely $\Vert\mathbf{c}\Vert$ and $\psi=\tan^{-1}(c_1/c_2)$, and integrating \eqref{fouriereq} with respect to $\psi$ (cf. \citet{Bartlett1964}). As a result, we have  
\begin{equation*}
f_{ii}(\varpi)=\lambda+ 2\pi \int_0^\infty \Vert \mathbf{c}\Vert\gamma_{ii}(\Vert\mathbf{c}\Vert)J_0(\Vert\mathbf{c}\Vert\varpi)d\Vert\mathbf{c}\Vert
\end{equation*}
where $J_0$ is the unmodified Bessel function of first kind of order zero as described in \citet{Watson1944} and $\varpi=\sqrt{w_1^2+w_2^2}$. Thus, the sprectrum depends on $\boldsymbol{\omega}$ only through the frequency magnitude $\varpi$. 

For the cross-spectrum of a stationary and isotropic process we obtain a similar simplification resulting in the real-valued function $C_{ij}(\boldsymbol{\omega})$. 

\subsection{Spatial dependence graph model}

We now define the spatial dependence graph model. Here, the underlying idea is to relate the structure of an undirected graph to the partial interrelation structure of a multivariate spatial point pattern. 

For an observed multivariate spatial point process, we identify the vertices of an undirected graph with the components of a multivariate spatial counting process. Then, two vertices $v_i$ and $v_j$ are not joined by an edge if and only if the component processes $N_i$ and $N_j$ are conditional orthogonal after elimination of the linear effect of $N_{\V\setminus\lbrace i,j\rbrace}$. As previously discussed, $N_i$ and $N_j$ are said to be conditionally orthogonal if and only if the partial spectral coherence vanishes at all frequencies $\boldsymbol{\omega}$. This is the case if the partial cross-spectrum $f_{ij\given\V\backslash\lbrace i,j\rbrace}(\boldsymbol{\omega})$, the inverse $g_{ij}(\boldsymbol{\omega})$ or equivalently the absolute rescaled inverse $d_{ij}(\boldsymbol{\omega})$ is zero at all frequencies $\boldsymbol{\omega}$.   

This leads to the following definition of a SDGM.
\begin{definition}
Let $\mathbf{N}_\V$ be a multivariate spatial counting process in $\mathbf{S}\subset\mathbbm{R}^2$. A spatial dependence graph model is an undirected graphical model $\G=(\V,\E)$ in which any $v_i\in\V(\G)$ encodes a component of $\mathbf{N}_\V$ and $\E(\G)=\lbrace (v_i,v_j): R_{i,j\given \V\backslash\lbrace i,j\rbrace}(\boldsymbol{\omega})\neq 0\rbrace$ such that
\begin{equation*}
\condindep{\lbrace N_i\rbrace}{\lbrace N_j\rbrace}{\lbrace N_{\V\backslash\lbrace i,j\rbrace}\rbrace}\Leftrightarrow (v_i,v_j)\notin\E(\G).
\end{equation*}
\end{definition}

Hence, a SDGM is a simple undirected graph in which conditional orthogonality relations can be read of missing edges. Precisely, two counting processes $N_i$ and $N_j$ are conditional orthogonal at all spatial lags after extracting the linear effect of all remaining components if the unordered pair $(v_i, v_j), ~i\neq j$ is not in $\E(\G)$. 

Additionally, further information can be obtained from the graph structure. The statement $\condindep{\lbrace N_i\rbrace}{\lbrace N_j\rbrace}{\lbrace N_{\V\backslash\lbrace i,j\rbrace}\rbrace}$ imposes that $\lbrace N_{\V\backslash\lbrace i,j\rbrace}\rbrace$ is a separator which intersects all paths from $\lbrace N_i\rbrace$ to $\lbrace N_j\rbrace$. Consequently, we have that $\lbrace N_i\rbrace$ and $\lbrace N_j\rbrace$ are not in the same component in $\G\setminus \lbrace N_{\V\backslash\lbrace i,j\rbrace}\rbrace$.

Extensions two disjoint subsets of $\mathbf{N}_\V$ are achieved naturally. 

\subsection{Markov properties of spatial dependence graph models}

This section discusses Markov properties with respect to spatial dependence graph models. As for ordinary undirected graphical models, these are the pairwise, local and global Markov property \citep{Lauritzen1996}. The global Markov property links conditional orthogonality statements to the graph theoretical notion of separation, while the pairwise and the local Markov properties are related to statements with respect to pairs of vertices or neighbourhoods of a vertices. 

To show that a SDGM is global Markovian, we assume that the spectral matrix is regular everywhere.  As previously discussed and similarly to \citet{Dahlhaus2000}, for four disjoint subprocesses $N_A$ to $N_D$ of $\mathbf{N}_\V$ we have that $\condindep{N_A}{(N_B, N_C)}{N_D}$ also implies $\condindep{N_A}{N_B}{N_D}$. Additionally, $\condindep{N_A}{(N_B, N_C)}{N_D}$ only holds if $\condindep{N_A}{N_B}{(N_C, N_D)}$ and $\condindep{N_A}{N_C}{(N_B, N_D)}$ are satisfied. Then, the SDGM is global Markovian, if for disjoint subsets of $\mathbf{N}_\V$ all paths from $N_A$ to $N_B$ are separated by $N_C$ or equivalently if $N_A$ and $N_B$ are not in the same component in $\G\setminus N_C$. The proof proceeds exactly the same way as in \cite{Lauritzen1996} and \cite{Dahlhaus2000}. 
 
\subsection{Estimation of spectral densities}

In this section we concern the estimation of auto- and cross-spectra from empirical data, namely the auto- and cross-periodograms. Assume we have observed a $d$-variate spatial point pattern within a rectangular region $\mathbf{S}\subset\mathbbm{R}^2$ with sides of length $l_x$ and $l_y$. Let $\lbrace\mathbf{s}_i\rbrace=\lbrace(x_i,y_i)\rbrace, i=1,\ldots,N_i$ denote the locations of events of type $i$. Respectively, $\lbrace\mathbf{s}_j\rbrace$ are the locations of events of type $j$. The auto- and cross-periodograms result from a discrete Fourier transform (DFT) of the locations $\lbrace\mathbf{s}_i\rbrace$ and $\lbrace\mathbf{s}_j\rbrace$.  

In detail, the DFT for events of type $i$ is given as  
\begin{align*}
F_i(p,q)&=(l_x, l_y)^{- \frac{1}{2}}\sum_{i=1}^{N_i}\exp\left(- 2\pi\iota N_i^{-1}\left(px_i + qy_i\right)\right)\\
&= A_i(p,q)+\iota B_i(p,q)
\end{align*}
where $p=0,1,2,\ldots$ and $q=0,\pm 1,\pm 2,\ldots$. From this expression, we obtain the auto-periodogram for frequencies $\boldsymbol{\omega}=(2\pi p/N,2\pi q/N)$ as 
\begin{eqnarray}\label{eq:autoperiodogram}
\hat{f}_{ii}(\boldsymbol{\omega})&=&F_i(p,q)\bar{F}_i(p,q)\\
&=& \lbrace A_i(p,q)\rbrace^2+\lbrace B_i(p,q)\rbrace^2.\nonumber
\end{eqnarray}
Here, $\bar{F}_i$ denotes the complex conjugate of $F_i$. 

Again we have that $\hat{f}_{ii}(\boldsymbol{\omega})=\hat{f}_{ii}(-\boldsymbol{\omega})$ such that it suffices to compute the periodogram for $p = 0, 1, \ldots, 16$ and $q = - 16, \ldots, 15$. Then, the maximum frequency amplitude of $\hat{f}_{ii}(\boldsymbol{\omega})$ is   
$\boldsymbol{\omega}_{max} = \sqrt{(32\pi/l_x)^2 + (32\pi/l_y)^2}$ (cf. \citet{Renshaw1983, Mugglestone1996a}).

The cross-periodogram can be calculated similarly. Thus, for frequencies $\boldsymbol{\omega}=(2\pi p/N,2\pi q/N)$ we have   
\begin{equation}\label{eq:crossperiodogram}
\hat{f}_{ij}(\boldsymbol{\omega})=F_i(p,q)\bar{F}_j(p,q).
\end{equation}

However, to omit bias in \eqref{eq:autoperiodogram} and \eqref{eq:crossperiodogram} at low frequencies, $\lbrace\mathbf{s}_i\rbrace$ and $\lbrace\mathbf{s}_j\rbrace$ are usually standardized or rescaled to the unit square prior to analysis (cf. \citet{Bartlett1964, Mugglestone1996a}). 
Then, assuming that the locations have been scaled to unit square, the DFT for events of type $i$ reduces to 
\[
F_i(p,q)=\sum^{N_i}_{i=1}\exp(-2\pi\iota(px_i+qy_i)).
\]

Finally, the co- and quadrature-spectrum are obtained by decomposing the cross-periodogram into the real and the imaginary part, namely
\[
\hat{C}_{ij}(\boldsymbol{\omega})=A_i(p,q)A_j(p,q)+B_i(p,q)B_j(p,q)
\]
and
\[
\hat{Q}_{ij}(\boldsymbol{\omega})=B_i(p,q)A_j(p,q)- A_i(p,q)B_j(p,q).
\]

\section{Applications to forest and crime data}\label{sec:2}

To illustrate the SDGM as introduced in this paper we discuss two real data examples taken from forestry and criminology. 

\subsection{Lansing Woods data}

\begin{figure}
\centering
\makebox{\includegraphics[scale=0.5]{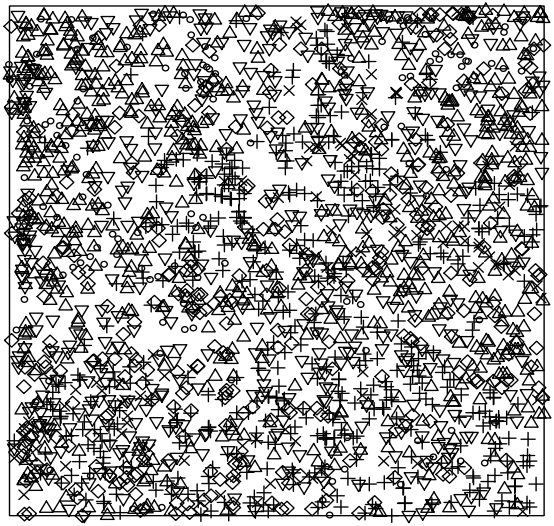}}  
\caption{
\label{fig:lansingData} Distribution of black oaks ($\circ$), hickories ($\Delta$), maples ($+$), miscellaneous trees ($\times$),  red oaks ($\diamond$) and white oaks ($\nabla$) in  Lansing Woods, Clinton County, Michigan USA}
\end{figure}

As a first example, we consider the distribution of forest stands for $2250$ trees in Lansing Woods, Clinton County, Michigan USA investigated by \citet{Gerrard1969}. This well-known data set has been analysed and discussed by several authors. In detail, the data set contains the locations and classifications of six different botanic tree species recorded in a $924$ft $\times$ $924$ft observation window, namely black oaks ($n=135$), hickories ($n=702$), maples ($n=514$), miscellaneous trees ($n=105$), red oaks ($n=346$) and white oaks ($n=448$). The locations of all species are shown in Figure \ref{fig:lansingData}.

One possibility to analyse such patterns is to treat each tree species separately and to calculate uni- and bivariate first- and second-order statistics.  For a detailed description of the first-order and second-order statistics we refer the interested reader to \citet{baddeley:rubak:tuner:15} and also \citet{Diggle2013}.  However, the second-order statistics only provide information on the pairwise structural interrelation between components processes. One possibility to overcome this limitation is to reduce the dimensionality of the
data and analyse cluster or latent structures of spatial point patterns. For example, \citet{Illian2006} applied a functional principal component analysis to second-order statistics in order to detect similarly behaving groups of component processes in the context of ecological plant communities. However, such dimensionality reduction techniques do not provide any information on the global interrelation structure between all component processes. 

In contrast, our objective of interest is to describe the global interrelations between all six tree species in space by means of a spatial dependence graph model. More precisely, our focus lies on the conditional relationship of two component processes which remain after elimination of all other alternative processes. 

\begin{figure}
\centering
\makebox{\includegraphics[scale=0.75]{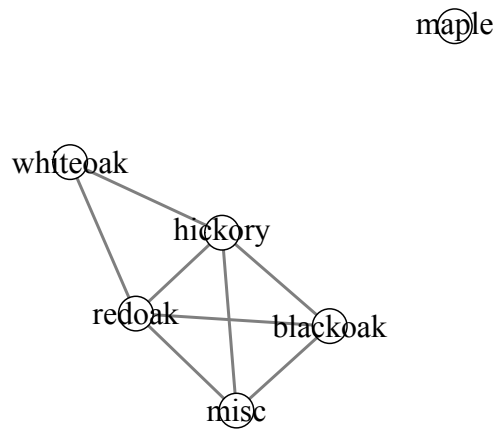}}
      \caption{\label{fig:lansingGM} Spatial dependence graph model for $6$ botanic tree species recorded in Lansing Woods}
\end{figure}

To obtain the SDGM, we calculated and inverted separate spectral matrices of dimension $6\times 6$ for all frequencies $\boldsymbol{\omega}$ for $p = 0, 1, \ldots, 16$ and $q = - 16, \ldots, 15$. Edges were drawn for all absolute rescaled inverse values whose supremum was above a given threshold $\alpha$  for all frequencies $\boldsymbol{\omega}$. For illustration, we set $\alpha=0.4$. Thus, edges indicate that the strength of the linear partial interrelation between two component processes is greater than $0.4$. The resulting SDGM is depicted in in Figure \ref{fig:lansingGM}. 

Interestingly, we observe that black oaks, hickories, miscellaneous trees, red oaks and white oaks form a subgraph while maple is not joined to any other tree species. From this, we can conclude that the spatial distribution of maples is not related to any other botanic species. In addition, two substructures can be found in the subgraph. On the one hand, hickories, red oaks and white oaks form a triangle. This indicates, that all three tree species are interdependent. On the other hand, a similar relation can be found for black oaks, hickories, red oaks and miscellaneous trees. Again, all these four tree species are conditionally interdependent. Finally, all paths from white oaks are separated by hickories as well as read oaks. From this we conclude that the spatial point pattern of white oaks is conditional orthogonal to all remaining tree species given the spatial point pattern of hickories and red oaks.       

\subsection{London crime data}

\begin{figure}
\centering
\makebox{\includegraphics[scale=0.4]{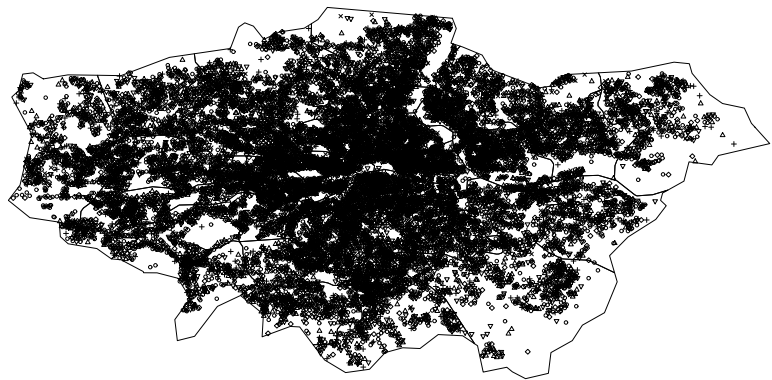}} \caption{\label{fig:londonData} Subset of six crime categories committed in the City of London in April 2015 recorded by the Metropolitan Police: anti-social behaviour ($\circ$), burglary ($\Delta$), violence and sexual offences ($\nabla$), vehicle crime ($\diamond$), drugs ($\times$), criminal damage and arson ($\diamond$)}
\end{figure}

As second illustration of the SDGM, we consider crime data provided by the British Home Office for the City of London which has been made available under the Open Government Licence. The open data has been downloaded from {\url{http://data.police.uk/data/}} and contains pairs of coordinates for different crime categories at street-level, either within a 1 mile radius of a single point or within a custom area of a street. The crime categories were generated by local officials. For our analysis we pre-selected a subset of $14$ crime categories which were reported within a one-month period in April $2015$. In total, $29914$ events were taken into account, duplicates excluded. In detail, the $14$ crime categories are anti-social behaviour $(n=10310)$, bicycle theft $(n=905)$, burglary $(n=3415)$, criminal damage and arson $(n=2668)$, drugs $(n=1065)$, public order $(n=1009)$, possession of weapons $(n=82)$, robbery $(n=557)$, shoplifting $(n= 362)$, theft from the person $(n= 543)$,  vehicle crime $(n=2825)$, violence and sexual offences $(n=3263)$, other thefts $(n=2791)$ and other crimes $(n=119)$.  
The distribution for a subset of $6$ crime categories is depicted in Figure \ref{fig:londonData}.

\begin{figure}
\centering
\makebox{\includegraphics[scale=0.5]{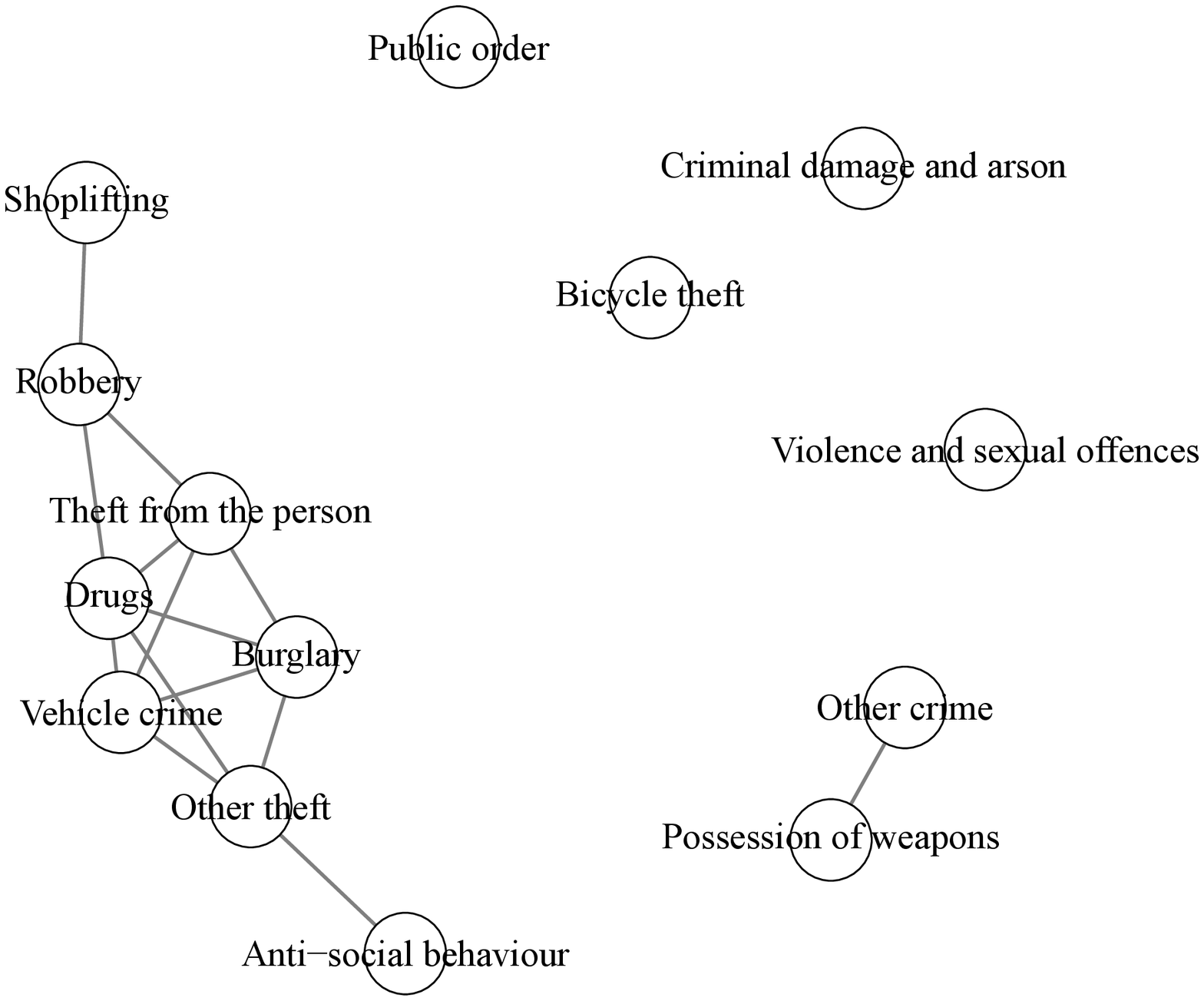}}
      \caption{\label{fig:LondonGM} Spatial dependence graph model for $14$ crime categories recorded within a one-month period  in April 2015 in the City of London.}
\end{figure}

As before, we computed and inverted separate spectral matrices of dimension $14\times 14$ for all frequencies $\boldsymbol{\omega}$ for $p = 0, 1, \ldots, 16$ and $q = - 16, \ldots, 15$. Again, we chose a threshold of $\alpha=0.4$. Hence, edges are missing if the supremum of the absolute rescaled inverse for all frequencies is less or equal to $0.4$. The resulting SDGM is depicted in Figure \ref{fig:LondonGM}.      

In this graph, we observe 4 isolated nodes (public order, bicycle theft, criminal damage and arson, violence and sexual offences),  a two-node subgraph (other crime and possession to weapons) and a subgraph with $8$ nodes (shoplifting, robbery, drugs, theft from the person, vehicle crime, burglary, other theft, anti-social behaviour). For the isolated nodes, one can conclude that all $4$ observed spatial point patterns are not interrelated to any other point pattern taken into account. In contrast, the $8$ node subgraph collects most crime categories related to theft offences. Here, anti-social behaviour and shoplifting are orthogonal conditional on all remaining $6$ crime categories included in this subgraph. In addition, the spatial pattern of shoplifting only depends on the observed pattern of robbery. Thus, conditional on the point pattern of robbery, none of the alternative crime categories contribute any additional information on the distribution of shoplifting.

Besides, various interesting substructures can be detected. Firstly, we see two triangle structures, namely a) robbery, drugs, theft from the person and b) burglary, vehicle crime, other theft. These triangles indicate that the structures of all three point patterns are interdependent. Besides, triangle structures also exist for example between the nodes other theft, vehicle crime and drugs. Secondly, we observe a complete subgraph consisting of drugs, theft from the person, vehicle crime, burglary. As before, this indicates an interdependence between the structures of all four spatial patterns.           

\section{Conclusion}\label{sec:3}
In this paper, we have introduced a novel graphical model which allows for the simultaneous exploration of global interrelation structures that emerge from multivariate spatial point processes. In the spatial dependence graph model, vertices represent components of a multivariate process and missing edges encode partial orthogonality relations. Thus, the SDGM presents a comprehensive picture of the global interrelations which exist in possibly high-dimensional spatial point patterns and is not affected by the number of observations taken into account.  

The spatial dependence graph model provides additional information to uni- and bivariate statistics and dimensionality reduction techniques and offers new insights into the conditional orthogonality structures between component processes. The definition of the SDGM is based on the  partial spectral coherence, which allows to differentiate between direct and induced effects.  Moreover, the SDGM is a nonparametric alternative to well-established distance-related measures which decomposes the complete covariance function into a sum of sines and cosines. 

The examples presented in the paper have been taken from forestry and criminology. For both data sets, the SGDM has detected several interesting structures. 

\bibliographystyle{rss}
\bibliography{spatgraph}

\end{document}